# Optimization of criteria for an efficient screening of new thermoelectric compounds: the TiNiSi structure prototype as a case-study

C. Barreteau[1], J-C. Crivello[1], J-M. Joubert[1] and E. Alleno[1]

[1]Univ Paris Est Créteil, CNRS, ICMPE, UMR7182, F-94320, Thiais, France

**Abstract**

High-throughput calculations are a very promising tool for screening a large number of compounds in order to discover new useful materials. Ternary intermetallic are thus investigated in the present work to find new compounds potentially interesting for thermoelectric applications. The screening of the stable non-metallic compounds required for such applications is obtained by calculating their electronic structure by DFT methods. In a first part, the study of the density of states at the Fermi level of well-known chemical elements and binary compounds allows to empirically optimize the selection criteria between metals and non-metals. In a second part, the TiNiSi structure-type is used as a case-study through the investigation of 570 possible compositions. This screening method leads to the selection of 12 possible semiconductors. For these selected compounds, their Seebeck coefficient and their lattice thermal conductivity are calculated in order to identify the most interesting one. TiNiSi, TaNiP or HfCoP could thus be compounds worth an experimental investigation.

**Introduction**

Thermoelectricity is an interesting technological path in the quest of alternative energies in order to reduce the energy consumption by recycling waste heat. The conversion efficiency of thermoelectric devices is related to the dimensionless figure of merit *ZT* of their constituting materials, which is defined as

$$ZT = S^2\sigma T/\lambda$$

with *S* the Seebeck coefficient, *σ* the electrical conductivity, *λ* the thermal conductivity and *T* the absolute temperature. However, even though efficient thermoelectric devices exist, they are limited to niche market because of their quite high cost compared to their yield of conversion. One way to improve this technology is to find new materials based on more abundant, less toxic and less expensive elements. Regarding these limitations, we decided to define a set of elements in which we can search for new interesting compounds (1). In this set of elements, hundreds of compounds have already been reported in crystallographic databases and even if some of them are known to be good thermoelectric compounds, there is still a very large number of possibly unknown compounds. In order to screen more efficiently those possibilities, we use high-throughput calculations. We will present an efficient method to screen ternary intermetallic compounds in order to find new promising thermoelectric compounds. In this article, we will focus on the equimolar compounds and use the orthorhombic TiNiSi structure type as a case-study. We choose the TiNiSi structure-type because even though it is a large family with more than 70 compounds already reported in the crystallographic databases (2, 3), it is the object of much less systematic studies when compared to other equimolar structure-type. For example, high-throughput calculations and screening have already been applied in several articles to equimolar compounds crystallizing in the half-Heusler family with very promising results (4-7).

Before possibly estimating the thermoelectric properties of the TiNiSi family, the main goal of the present work is to define an efficient method to discriminate among the investigated compounds those which may be stable and semiconducting. In our previous investigation (1), we compared the enthalpy of formation of a same combination of elements in four structure types to determine the stable one. In the present work, for the compounds stable in the TiNiSi structure-type, investigation on the electronic band structure will be carried out in order to exclude the metallic compounds and keep only the possible semiconductors and semimetals for thermoelectric application. A definition and the application of our criterion of exclusion will be presented in the present article. Finally, for the most interesting compounds, the Seebeck coefficient and the thermal conductivity will be estimated from calculations based on the BoltzTrap package (8, 9) and the Slack equation (10, 11) respectively. All the results will be compared with the literature and the selected compounds will be thus sorted between each other to highlight the most promising.

**Computational details and methodology**

In our previous work (1), a combinatorial approach has been implemented in order to screen all possible *T-M-X* combinations with the 1:1:1 stoichiometry generated within a restrained set of chemical elements. In this set of elements, *T* is a transition metal from the Ti, V, and Cr columns or Sr, Ba, Y and La, *M* an element from the first line of transition metals and *X* a *sp* element (Al, P, Si, Sn and Sb). All *T-M-X* combinations have been calculated in four different structure-types: the orthorhombic TiNiSi (*Pnma*), the cubic MgAgAs (*F-43m*, half-Heusler), the two hexagonal ZrNiAl (*P-62m*) and BeZrSi (*P*6$_3$/*mmc*). The 570 possible configurations are calculated in these 4 structure-types by systematically assigning a unique element from our *T-M-X* nomenclature to each crystallographic site, discarding "inverse" configurations or site mixing, thus yielding 2280 different compounds. Among the 570 unique configurations, 317 compounds present a negative enthalpy of formation in the orthorhombic structure.

The calculations of the stability and ground state properties are based on the density functional theory (DFT), which allows to obtain the enthalpy of formation at 0 K for each compound (by total energy difference with the elemental reference state) as well as its electronic structure. They were conducted using the projector augmented wave (PAW) method implemented in the Vienna *ab initio* Simulation Package (VASP) (12-15). The exchange correlation was described by the generalized gradient approximation modified by Perdew, Burke and Ernzerhof (GGA-PBE) (16). Energy bands up to a cut off energy E = 600 eV were used in all calculations. A high-density *k*-points meshing was employed for Brillouin zone integrations in the TiNiSi structure-type (7 × 11 × 6), MgAgAs (21 × 21 × 21), ZrNiAl (19 × 19 × 37) and BeZrSi (21 × 21 × 10). These parameters ensured good convergence for the total energy. The convergence tolerance for the calculations was selected as a difference on the total energy within $10^{-6}$ eV. We performed for each structure-type, volume and ionic relaxation steps and we considered magnetism for all the configurations. Blöchl correction was considered in a final step calculation (17). Phonon dispersion bands were obtained by computing the atomic forces for different, finite atomic displacements (18) and a subsequent calculation and integration over the corresponding phonon frequencies within the harmonic approximation, i.e. without any volume dependence, using the Phonopy code (19, 20). The phonon calculations were carried out within a 1 x 2 x 1 supercell (24 atoms) with 12 atomic displacements for the TiNiSi structure-type.

The Seebeck coefficient was calculated from the Boltzmann transport theory as implemented in the BoltzTrap simulation package (v2) (8, 9). The BoltzTrap code uses the electronic structure calculated from VASP as an input data and assumes a constant relaxation time for the electrons.

The first step of our methodology deals with the stability of the calculated compounds. Indeed, a positive enthalpy of formation leads to the exclusion of the combination as it is obvious that the ternary compound is unstable toward the decomposition into the pure elements. In a second step, the most negative enthalpy of formation of the four-competing structure-types at the same composition allows to select the most probable stable structure. Comparison with the literature and the crystallographic database has confirmed the robustness of our model (1).

Definition of the non-metallic criteria

In our methodology, in addition to the criterion on the stability of each compound, it is important to find a simple and robust criterion to exclude metallic compounds. For this purpose, we decided to focus our screening method on the density of states (DOS) at the Fermi level. It is of course well known that when the DOS is zero at the Fermi energy ($E_F$), a semiconducting or insulating ground state is expected. However, DFT-PBE is known (21, 22) to not only systematically underestimate the gap magnitude but even "turn semiconductors into bad or semi-metal". To compensate for this spurious effect, a non-zero value of the DOS at $E_F$ is thus required to assess an unknown ground state : not too strict for not missing interesting compounds and not too high for being efficiently selective. To estimate this value, preliminary calculations have been performed on the pure chemical elements in their stable structure at 0°K and on known semiconducting binary compounds. Indeed, most elements of the periodic table have been investigated, except the *4f* elements and the noble gas, whereas the studied binary compounds have been chosen among known semiconductors or semi-metals (InTe, GaAs or HgTe).

The calculated DOS are presented in Fig. 1. As expected, all the metallic elements (block *s, d*) have a non-zero DOS at $E_F$ consistent with their metallic ground-state. In the group ***III***, (B, Al, Ga, In), the last three elements present a non-zero DOS at $E_F$ in agreement with their metallic ground state whereas B has a zero DOS consistent with its semiconducting ground state (23). When looking at the first elements from the group ***IV*** (C, Si, Ge), it can easily be noticed that all the DOS at $E_F$ exhibit a zero value. The last two elements of this group, Sn and Pb, have non zero DOS at $E_F$, respectively 0.05 state eV$^{-1}$ by atom and 0.5 state eV$^{-1}$ by atom. This very low DOS at $E_F$ for Sn is in agreement with previous results (24) which have shown that "gray tin", the α form of Sn (*A*4), which is the stable allotrope at low temperature (25), is semi-conducting. Elemental Pb displays a DOS at $E_F$ similar to the metals of the block *d*. In the group ***V*** (P, As, Sb, Bi), P has a zero DOS at $E_F$ whereas As, Sb and Bi have a DOS smaller than 0.1 state eV$^{-1}$ by atom. This agrees with the semi-metallic ground state of these last three elements, while elemental P is a semiconductor as reported in reference (26). In the group ***VI*** with Se and Te the DOS at $E_F$ is zero as these elements are reported as semiconductors (27, 28). Finally, all the known binary semiconductors exhibit a zero DOS at $E_F$.

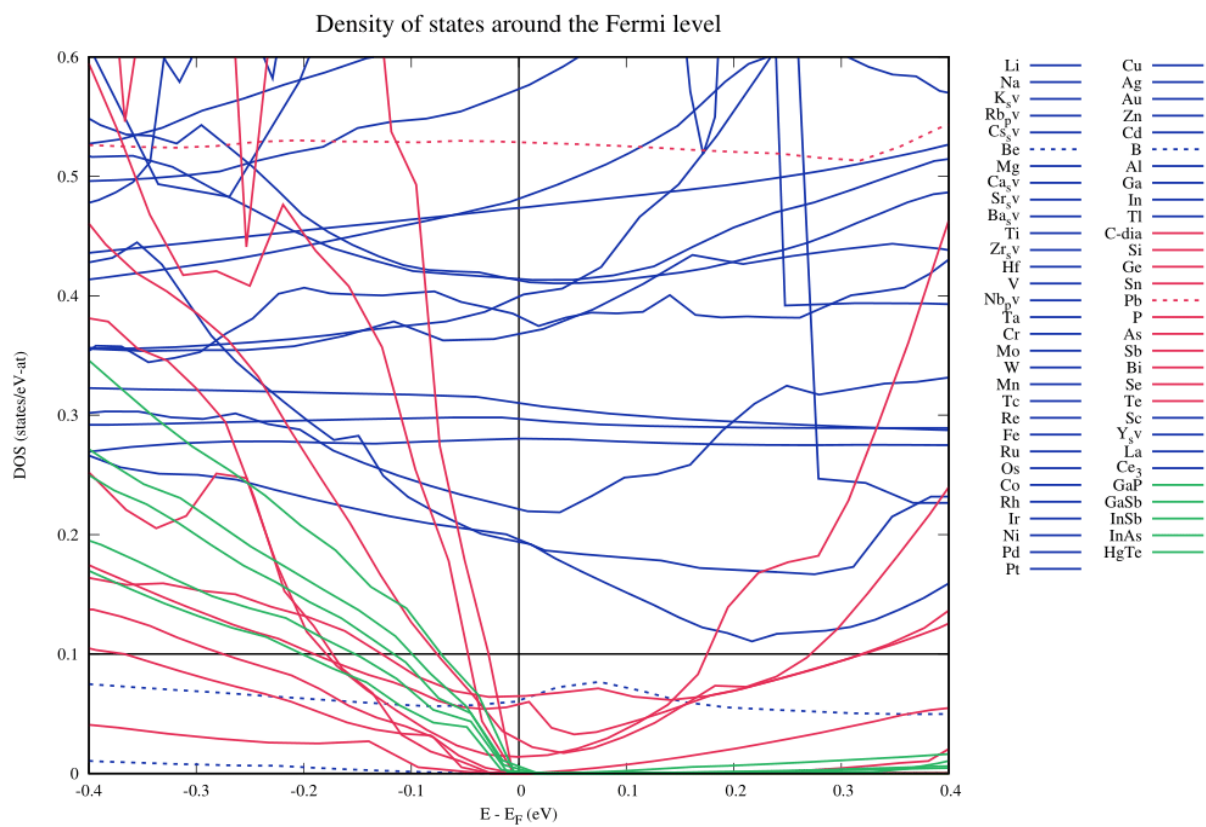

*Figure 1 : Evolution of the density of states (DOS) as a function of the energy for pure element and binary compounds; the Fermi level is chosen as the origin of the energy; binary compounds in green, elements from the group IV, V and VI which are semi-metallics or semiconductors in red and metals are in blue, dashed line corresponds to elements which differ from the other elements of their column (Pb, Be, B).*

In the Figure 1, a significant interval at $E_F$ can be noticed between the DOS of Sn or As and these of Ga and Zn, two metallic elements. Thus, with a selection limit set at 0.1 state eV$^{-1}$ by atom for the DOS at $E_F$, semiconductors as well as elements with a semi-metallic ground state such as As or Sb are kept whereas metals are excluded. Only one exception to this criterion is found: with this limit, Be, which is known as a metal is also selected. However, Be is known to exhibit a very low DOS at $E_F$, despite its metallic behaviour (29-32). Overall, this limit ensures a judicious selection with a very small number of ambiguous results. It is nonetheless essential to keep the ambivalent results as the calculations method is based on the GGA-PBE approximations known to systematically underestimate the gap.

**Twelve selected compounds in the TiNiSi structure type**

Among the 317 compounds more stable in the TiNiSi structure-type than in the other structure type only 12 compounds are found as potential non-metallic, according to the electronic criterion set at 0.1 state eV$^{-1}$ by atom for the DOS at $E_F$, and reported in Table 1. All of them present a mechanical stability since no imaginary frequency is observed in the phonon bands. We will first review the existence and structure of each selected compound. Then for each candidate, a more specific study of the electronic structure will be presented. Finally, in order to assess their thermoelectric properties, calculated values of the Seebeck coefficient as well as of the thermal conductivity will be examined.

*Table 1 : Selection of non-metallic compounds in the TiNiSi structure type*

| *TMX* Compounds | Crystal databases | Enthalpy of formation (kJ mol$^{-1}$) | DOS at the Fermi level (state. eV$^{-1}$ by atom) |
|---|---|---|---|
| **TiCuSb** | Unreported | -20.90 | 0.092 |
| **ZrZnSi** | Unreported | -62.22 | 0.098 |
| **HfMnSb** | Unreported | -26.15 | 0.098 |
| **HfCuSb** | Unreported | -27.12 | 0.015 |
| **HfZnSi** | Unreported | -52.84 | 0.098 |
| **TaCuSi** | Unreported | -44.15 | 0.043 |
| **MoZnSi** | Unreported | -15.53 | 0.062 |
| **TiMnSb** | BeZrSi | -23.06 | 0.071 |
| **TiCoP** | TiNiSi | -109.95 | 0.068 |
| **TiNiSi** | TiNiSi | -88.47 | 0.061 |
| **HfCoP** | TiNiSi | -109.99 | 0.032 |
| **TaNiP** | TiNiSi | -73.48 | 0.062 |

Over the 12 compounds, to our knowledge, 7 are not reported in crystal databases. However, a more cautious study of the literature, can in several cases, give supplementary information about the presumable existence of these compounds. For instance, when they are published, ternary phase diagrams provide additional information. Thus, for TiCuSb, ZrZnSi, HfCuSb and TaCuSi, there is no equimolar compound reported in the ternary phase diagram (33-37). In these cases, it is unlikely that an equimolar compound is stable in the ternary orthorhombic structure but it cannot be formally discarded. For the 3 other unreported compounds, HfMnSb, MoZnSi and HfZnSi, no additional information or ternary phase diagram are available, it is then difficult to conclude on their existence.

Among the 5 reported compounds, TiMnSb has a reported structure in the literature (BeZrSi structure type) (38) different from the TiNiSi prototype which should be more stable according to our calculation. However, as discussed in our previous work (1), this disagreement can be explained by the experimental conditions for the synthesis. Indeed, Noda *et al.* (38) have obtained TiMnSb under extreme conditions (high pressure and high temperature) which differs drastically from those used for the present calculations.

Finally, 4 compounds are reported in the crystallographic databases in the orthorhombic structure. In the cases of TiCoP and TaNiP, the literature only deals with their crystallographic structure and there is no investigation of their electronic properties (39, 40). Additional information is provided for the other two cases: HfCoP and TiNiSi. Recently, Huang *et al.* (41) investigated the thermoelectric properties of TiNiSi-based solid solutions. Theoretical results, based on DFT calculations with PBE-GGA, confirmed the presence of a pseudo-gap in the electronic band structure of TiNiSi. Measurements of the electric conductivity versus the temperature confirmed that TiNiSi is a semimetal while thermoelectric properties were improved by the substitution of Ti by V (41). These results are in good agreement with a previous report by Kong *et al.* (42) who have also highlighted the semimetallic ground state of TiNiSi, thanks to their calculations. HfCoP, based on an extended Hückel calculation by Kleinke *et al.* (43), could be considered as a metal. However, neither electronic density of states nor electrical resistivity data are reported in this reference. The ground state of HfCoP is currently unknown.

**Seebeck coefficient and Lattice thermal conductivity at finite temperature**

The Seebeck coefficient was calculated at finite temperature within the semi-classical Boltzmann transport theory with the BoltzTrap package, assuming a constant relaxation time.

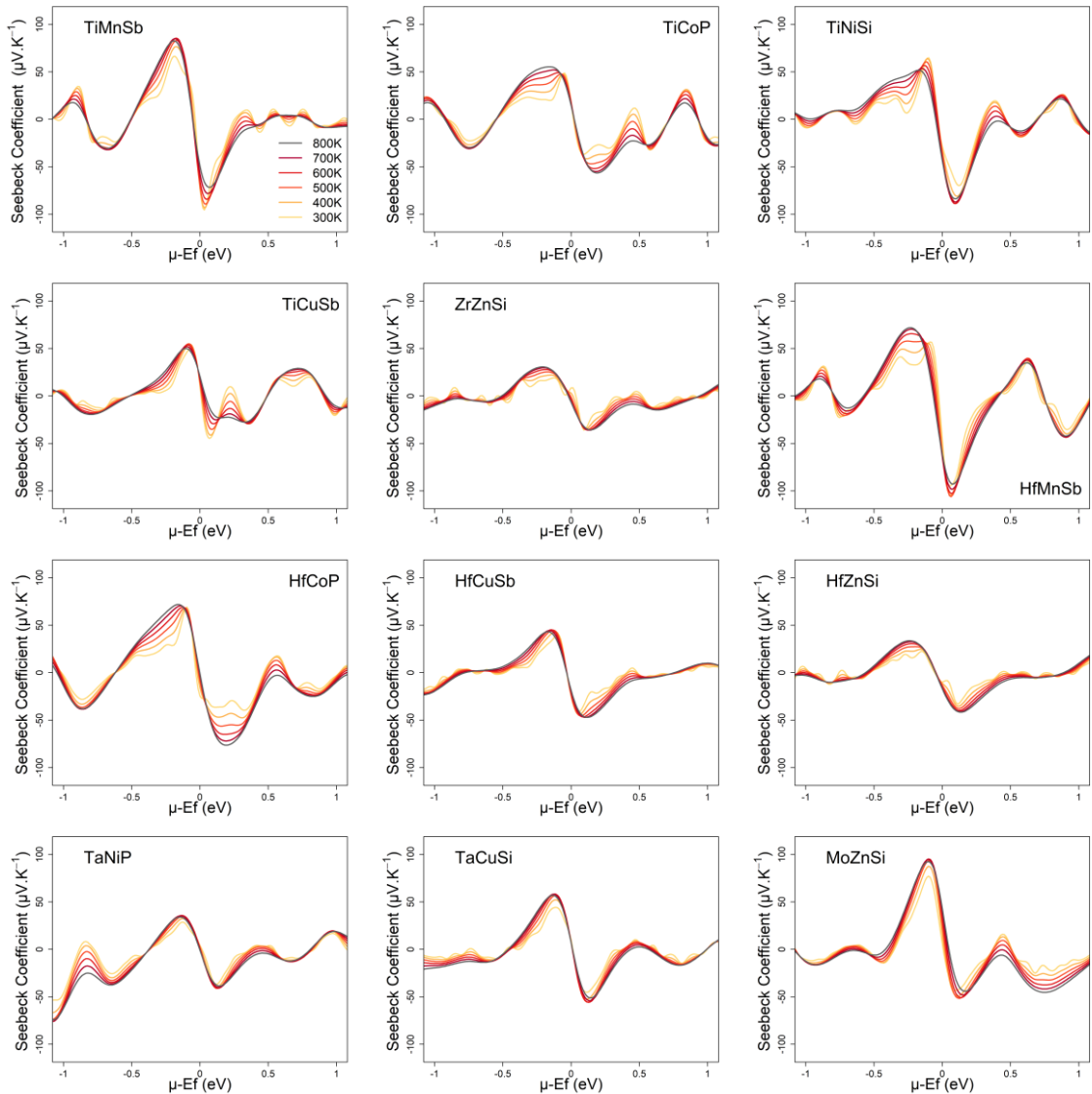

*Figure 2 : Seebeck coefficient as a function of the chemical potential for each compound in the temperature range 300 (orange) to 800K (black).*

The Seebeck coefficient has been plotted as a function of the chemical potential ($\mu$) for each selected compound. Figure 2 presents the variations of *S* at several temperatures between 300K and 800K for each compound.

Several compounds are less interesting than others as they do not reach at any value of µ or at any temperature, values larger than +-50 µV K$^{-1}$. Indeed, ZrZnSi or HfZnSi do not seem very promising regarding their quite low Seebeck coefficient contrary to TiMnSb or MoZnSi for example. For a simpler analysis, maximum negative and positive values of Seebeck coefficient in the vicinity of the Fermi energy, [-0.5 to 0.5 µ- $E_F$ (eV)], are listed in the Table 2. In this table, values at 700K are reported, in order to compare the compounds in the same way as it will be done for the lattice thermal conductivity. Besides, we choose to focus on the maximal values of Seebeck coefficient when µ is near the Fermi level for easy comparison.

Table 2 : Maximum of the calculated Seebeck coefficient in the vicinity of the Fermi level and calculated lattice thermal conductivity at 700K for the TiNiSi compounds

| TMX compounds | $S_{max}$ (µV K$^{-1}$) > 0 (700K) | $S_{max}$ (µV K$^{-1}$) < 0 (700K) | Calculated Lattice Thermal conductivity at 700K (W m$^{-1}$ K$^{-1}$) |
|---|---|---|---|
| TiCoP | 52 | -55 | 15.7 |
| TiNiSi | 54 | -87 | 13.7 |
| HfCoP | 71 | -72 | 11.6 |
| TaNiP | 41 | -76 | 7.6 |
| TiMnSb | 85 | -78 | 5.4 |
| HfMnSb | 70 | -98 | 6.1 |
| HfZnSi | 33 | -41 | 7.4 |
| MoZnSi | 95 | -48 | 6.5 |
| TiCuSb | 52 | -28 | 3.9 |
| ZrZnSi | 30 | -36 | 9.9 |
| HfCuSb | 44 | -47 | 3.7 |
| TaCuSi | 58 | -54 | 9.7 |

As mentioned before, ZrNiSi, HfZnSi, HfCuSb and TaCuSi present quite low Seebeck coefficient near the Fermi energy, regardless their conduction type. Considering both positive and negative values, HfMnSb can be considered as the best compound when other compounds could be promising at most in one of the two conduction types. For example, TiNiSi has a negative Seebeck coefficient twice as large as its the positive counterpart whereas it is the opposite for MoZnSi. No tendencies can be noticed in the evolution of the Seebeck coefficient regarding the nature of each *T*, *M* or *X* element in these compounds.

Lattice thermal conductivity, which is essential in the determination of ZT, can be estimated by the Slack equation (10, 11) by assuming that heat is conducted only by acoustic phonons (44, 45), through the equation:

$$K_L = 3.1 \times 10^{-6} \frac{M \, \theta_D^3 \delta}{\gamma^2 n^{\frac{2}{3}} T}$$

where $\theta_D$ is the Debye temperature in K, $\delta^3$ is the volume per atom in Å$^3$, *n* is the number of atoms in the primitive unit cell, $M_{av}$ is the average atomic mass in atomic unit and *γ* is the average Grüneisen parameter. The Debye temperature and the Grüneisen parameter are evaluated from the transverse and longitudinal sound velocity which are obtained from the calculated elastic modulus. The calculated elastic properties have been obtained from the phonon calculations for each potential compound (see SI).

To fulfil the condition that the temperature *T* is higher than the Debye temperature $\theta_D$ for all the compounds, the values presented in Table 2 were calculated at 700K.

The range of the lattice thermal conductivity is quite broad as the largest value is 5 times larger than the lowest, with values between 3 and 15 W m$^{-1}$ K$^{-1}$. These values are quite large compared to the best thermoelectric materials as $Bi_2Te_3$ or $Sb_2Te_3$ which exhibit low thermal lattice thermal conductivity (~ 2 W m$^{-1}$ K$^{-1}$). They are nonetheless still promising because they were calculated for a perfect crystal and it is known that defects or impurities can strongly decrease the lattice thermal conductivity. For instance, the half Heusler HfNiSn displays a thermal conductivity of 10 W m$^{-1}$ K$^{-1}$ at 300K (46) while the solid solution $Hf_{0.75}Zr_{0.25}NiSn_{0.99}Sb_{0.02}$ displays a thermal conductivity of 5 W m$^{-1}$ K$^{-1}$ at 300K and ZT = 0.9 at 800K (47).

Keeping the **X**-element constant and considering the **T** element, as expected, the lattice thermal conductivity is in general lower for "heavier" **T** in the same columns of the periodic table as for **T**CuSb or **T**CoP. In the case of **T**MnSb, the trend is reversed as $\kappa_L$(TiMnSb) is lower than $\kappa_L$(HfMnSb). Keeping the T element constant and considering the X element, it can be noticed that the antimonides present the lowest thermal conductivity compared to the silicides or the phosphides, in consistency with the larger atomic mass of antimony.

**Discussion**

Keeping in mind that the values presented here are from calculations based on DFT-PBE and constant relaxation time approximation implemented in a perfect crystal, it nonetheless offers the possibility to identify the most promising compounds among the 12 selected ones. For the last four compounds in Table 2, even though, lattice thermal conductivity is quite low for both **T**CuSb compounds, their Seebeck coefficients are always quite low. If one adds to these low values the fact that, according to the reported ternary phase diagram, they are not stable, it is probably uninteresting to synthetize them.

On the other hand, HfMnSb and MoZnSi combine a large Seebeck coefficients and a small lattice thermal conductivity that could lead to interesting properties. Nonetheless, their existence is not established.

Regarding the four compounds which have already been reported in the literature, a ranking is also possible. The highest value of Seebeck coefficient is reached in TiNiSi, which has already been reported as interesting for its thermoelectric properties in the solid solution $Ti_{1-x}V_xNiSi$ (41). The presently calculated values of the Seebeck coefficient are in good agreement with the calculated values obtained by Huang *et al.* (41). However, the measured values (41) differ from the calculated one, but this difference can be explained by the charge carrier concentration of the pristine TiNiSi, which is around $2 \times 10^{21}$ cm$^{-3}$, this value indicates the presence of intrinsic defects in the sample. Then, HfCoP seems to be more interesting than TiCoP as its Seebeck coefficient is larger and associated with a lower lattice thermal conductivity. TaNiP exhibits the lowest lattice thermal conductivity of the four reported compounds and negative Seebeck coefficient of the same order of magnitude as in HfCoP.

**Conclusion**

After determining the stability of thousands of *TMX* compounds in 4 structure types, a more specific investigation has been made on the compounds which crystallize in the TiNiSi structure type. In a second stage screening potential thermoelectric compounds, the electronic band structure has been examined and an effective criterion of the DOS at the Fermi level has been established by comparing calculated DOS of known semiconducting elements and compounds. Then, this empirical criterion of 0.1 state eV$^{-1}$ by atom has been applied to the TMX compounds stable in the orthorhombic structure and has led to the selection of 12 non-metallic compounds among the 317 possible compounds. Additional phonon and BoltzTrap calculations allowed to rank the 12 selected compounds in order to identify those which are the most interesting to synthetize from the point of view of their anticipated thermoelectric properties. Although MoZnSi could be an interesting compound as it presents the highest Seebeck coefficient associated to a small lattice thermal conductivity, there is no information about its possible stability. So, it seems more relevant to focus on HfCoP and TaNiP as starting points. Indeed, both compounds are already reported in the literature and present the best compromise between high Seebeck coefficient and a low lattice thermal conductivity.

## Acknowledgements

DFT calculations were performed using HPC resources from GENCI-CINES (Grant A0060906175) and financial support from the Agence Nationale de la Recherche is acknowledged (ANR-18-CE05-0010-01).